\documentstyle[epsfig,prc,tighten,aps,preprint]{revtex}

\begin{document}

\title{Progress toward a new measurement of the parity violating
  asymmetry in $\vec{n}+p \rightarrow d+\gamma$}

\author{W.M. Snow,$^{(1)}$ W.S. Wilburn,$^{(2)}$ J.D. Bowman,$^{(2)}$
  M.B. Leuschner,$^{(3)}$ S.I. Penttil{\"a},$^{(2)}$ V.R.
  Pomeroy,$^{(3)}$ D.R. Rich,$^{(1)}$ E.I. Sharapov,$^{(4)}$\ and V.
  Yuan$^{(2)}$}

\address{$^{(1)}$ Department of Physics, Indiana University,
  Bloomington, Indiana 47405 \\
  $^{(2)}$ Los Alamos National Laboratory, Los Alamos, New Mexico 87545 \\
  $^{(3)}$ Department of Physics, University of New Hampshire, Durham, 
  New Hampshire 38245 \\
  $^{(4)}$ Joint Institute for Nuclear Research, 141980 Dubna, Russia \\}

  \date{\today}
  \maketitle
  \widetext
    
  \begin{abstract}
    We outline the motivation and conceptual design for a new
    experiment aimed at a 10-fold improvement in the accuracy of the
    parity-violating asymmetry $A_{\gamma}$ in the angular
    distribution of 2.2~MeV gamma rays from the $\vec{n}+p\rightarrow
    d+\gamma$ reaction. This observable is primarily sensitive to the
    weak pion-nucleon coupling $H_{\pi}^{1}$. A proof-of-principle
    experiment using unpolarized low-energy neutron capture on
    polyethylene and an array of 12 CsI detectors operated in current
    mode has been performed.  Results of this test experiment
    including the current mode signal, electronic noise and detector
    sensitivity to magnetic fields are reported.
  \end{abstract}
  
  \section{Motivation}
    \label{sec:the}
  
    The flavor-conserving quark-quark weak interaction remains the
    most poorly tested aspect of the electroweak theory. One can judge
    the slow rate of progress in our understanding of this sector by
    noting the similar state of affairs described in two reviews of
    the subject conducted a decade apart \cite{Ade85,Hae95}. The
    reasons for the slow advance are both theoretical and
    experimental. The experimental problems stem from the small size
    of weak amplitudes relative to strong amplitudes (typically
    $\approx10^{-7}$ at low energies). The theoretical difficulties
    are encountered in trying to relate the underlying electroweak
    currents to low-energy observables in the strongly interacting
    regime of QCD\@. The current approach is to split the problem into
    two parts. The first step is to map QCD to an effective theory
    expressed in terms of the important degrees of freedom of low
    energy QCD, mesons and nucleons. In this process, the effects of
    quark-quark weak currents appear as parity-violating
    meson-nucleon-nucleon couplings \cite{Des80}. The second step is
    to use this effective theory to calculate electroweak effects in
    the \emph{NN} interaction and to determine the weak couplings from
    experiments.
  
    A naive analysis of the structure of the quark-quark weak current
    implies that the isovector part of the weak current should be
    strongly dominated by the neutral current contribution. In terms
    of the meson-exchange picture of the weak \emph{NN} interaction,
    this means that the weak pion exchange is particularly
    interesting, since it should be dominated by neutral currents. It
    is also the longest range component of the weak \emph{NN}
    interaction, and therefore presumably the most reliably calculable
    in the \emph{NN} system.  Finally, it is worth pointing out that
    the exchange of neutral currents between quarks has never been
    isolated experimentally.  For all of these reasons, the coupling
    constant for the weak $\pi$ exchange, $H_{\pi}^{1}$, is of special
    interest\footnote{We have used the notation of Adelberger and
      Haxton \cite{Ade85}, where
      $H_{\pi}^{1}=F_{\pi}=g_{\pi}f_{\pi}/\sqrt{32},
      H_{\rho}^{1}=F_{1}=-g_{\rho}h_{\rho}^{1}/2,
      H_{\omega}^{1}=G_{1}= -g_{\omega}h_{\omega}^{1}/2~{\rm
        and}~H_{\rho}'^{1}=H_{1}= -g_{\rho}h_{\rho}'^{1}/4$.}.
    
    The size of $H_{\pi}^{1}$ is not known. The most reliable
    information on the strength of $H_{\pi}^{1}$ is believed to come
    from measurements of the circular polarization of 1081~keV gamma
    rays from $^{18}$F \cite{Pag87}. The current results have been
    interpreted as an upper limit of the weak pion-nucleon-nucleon
    coupling, $H_{\pi}^{1}\leq3.4\times10^{-7}$.  This value is
    $3\sigma$ smaller than the expected value from the theory,
    $H_{\pi}^{1}=10.8\times10^{-7}$ (DDH ``best value'' \cite{Des80}).
    This result has led to speculation that quark-quark neutral
    currents might be suppressed in $\Delta I=1$ processes. New
    information on $H_{\pi}^{1}$ comes from the recent observation of
    nuclear parity violation in an atomic parity-violation experiment
    in $^{133}$Cs \cite{Woo97}. This experiment detected for the first
    time the parity-violating nuclear anapole moment. Theoretical
    estimate for the value of $H_{\pi}^{1}$ inferred from this
    experiment, is $H_{\pi}^{1}=22.6\pm5.0({\rm exp.})\pm8.3({\rm
      theor.})  \times10^{-7}$ \cite{Fla97}, which is significantly
    larger than the upper limit set by the $^{18}$F experiments.  This
    disagreement is almost certainly due to problems in the nuclear
    structure calculations necessary to extract $H_{\pi}^{1}$ from
    these measurements, and there is speculation that $H_{\pi}^{1}$
    may be modified in the nuclear medium \cite{Des98}. At the same
    time, the DDH prediction for $H_{\pi}^{1}$ has sharpened to
    $0$--$6\times10^{-7}$ \cite{Des96} due to improved knowledge of
    the QCD coupling constant and the treatment of quark masses.
    
    An accurate measurement in the nucleon-nucleon system sensitive to
    $H_{\pi}^{1}$ is needed to resolve these inconsistencies. The
    \emph{NN} system is simple enough that the measured asymmetry can
    be related to the weak meson-nucleon-nucleon coupling with
    negligible uncertainty to nuclear structure.  Parity violation in
    the neutron-proton system is primarily sensitive to weak $\pi$-
    and $\rho$-exchange, which are the longest-range contributions. An
    analysis of the available low-energy channels in the
    neutron-proton system indicates \cite{Ade85} that
    parity-nonconserving (PNC) effects in the reaction
    $\vec{n}+p\rightarrow d+\gamma$ are almost entirely due to the
    weak pion exchange.
    
    The relationship between the PNC asymmetry $A_{\gamma}$ and
    $H_{\pi}^{1}$, where $A_{\gamma}$ is the correlation between the
    direction of emission of the gamma ray and the neutron
    polarization, is calculated to be \cite{Ade85}
    \begin{equation}
      A_{\gamma}=-0.045\left(H_{\pi}^{1}-0.02H_{\rho}^{1}
        +0.02H_{\omega}^{1}+0.04H_{\rho}^{'1}\right).
    \end{equation}
    This result is consistent on the coefficient of $H_{\pi}^{1}$ with
    an earlier calculation by Desplanques and Missimer \cite{Des78}.
    The final result from the last experiment \cite{Cav77} was
    $A_{\gamma}=-1.5\pm4.8\times10^{-8}$ \cite{Alb88}. This value is
    neither sensitive enough to address the inconsistencies in the size
    of $H_{\pi}^{1}$, or to reach the range of values for
    $H_{\pi}^{1}$ predicted by theories \cite{Des80,Hen96,Kap93}.
    
\begin{figure}
      \centering
      \epsfig{file=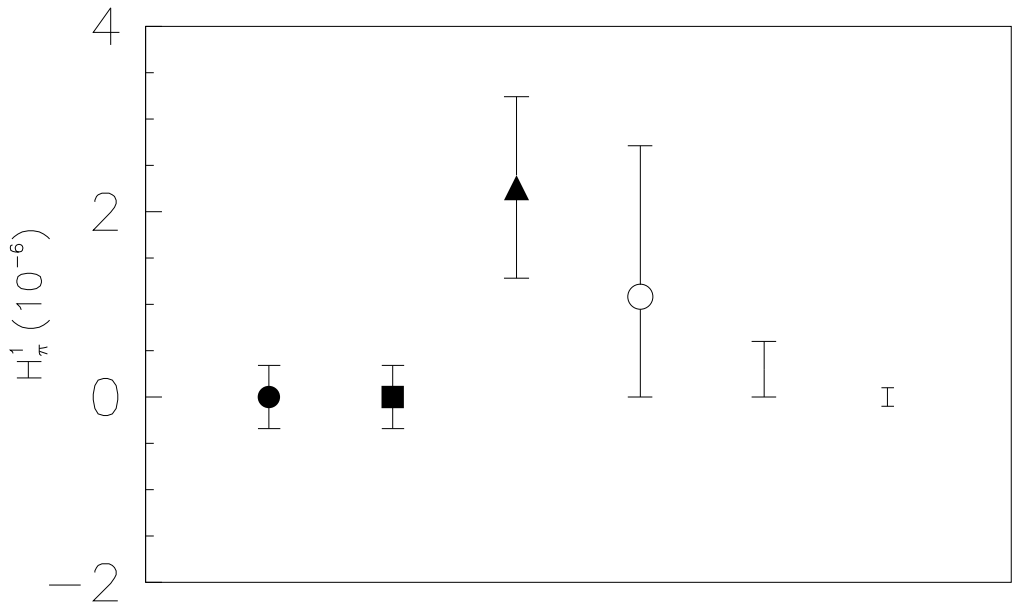,width=\textwidth}
      \caption{Values of $H_{\pi}^{1}$ from (left to right) the
        earlier $\vec{n}+p\rightarrow d+\gamma$ experiment, $^{18}$F
        experiments, $^{133}$Cs experiment, DDH theoretical estimate,
        and Desplanques theoretical estimate. The last value
        represents the expected uncertainty from the proposed
        experiment. References and explanation are found in the text.}
      \label{fig:hp0}
    \end{figure}

    In this paper, we discuss a new experiment to measure $A_{\gamma}$
    to a precision of $\pm5\times10^{-9}$ which will determine
    $H_{\pi}^{1}$ to $\pm1\times10^{-7}$ (see figure~\ref{fig:hp0}).
    Such a result will clearly distinguish between the $^{18}$F and
    $^{133}$Cs values for $H_{\pi}^{1}$ as well as between different
    predictions given by theories of the weak interaction of hadrons
    in the non-perturbative QCD regime.  There is also a strong
    possibility that a non-zero result will be seen and that the value
    of $H_{\pi}^{1}$ will finally be known.

  \section{Requirements for an Accurate Measurement of  $A_{\gamma}$ in
    the Reaction $\vec{n}+p\rightarrow d+\gamma$}
    \label{sec:exp}
  
    To determine $H_{\pi}^1$ with an uncertainty of
    $1.0\times10^{-7}$, 10\% of the DDH best value, we need a
    statistical uncertainty of $0.5\times10^{-8}$ on $A_{\gamma}$
    which means that we have to detect about $4\times10^{16}$ gammas
    from the $\vec{n}+p\rightarrow d+\gamma$ reaction. On the other
    hand, we have to keep the systematic errors below the statistical
    error.  Among the essential aspects of the proposed experiment
    which led us to the choice of issues to be addressed in the
    proof-of-principle experiment, are the following:
    \begin{enumerate}
    \item The number of events required to achieve sufficient
      statistical accuracy in a reasonable time immediately leads to
      the conclusion that the 2.2~MeV gamma rays must be counted in
      current mode.
    \item Neutron fluxes available at present neutron sources are not
      high enough to achieve the required statistical accuracy in a
      small amount of time.  Therefore, it is important to demonstrate
      that the electronic noise introduced by the current-mode
      measurement technique is negligible compared to the shot noise
      due to the discrete nature of the charge deposited by each gamma
      ray and number of photoelectrons produced by the neutron capture
      event.
    \item The tiny parity violating signal will be isolated by
      flipping the neutron spin, since the real asymmetry will change
      sign under spin reversal, while spin-independent false
      asymmetries will not. The neutron spin can be flipped by either
      static or RF magnetic fields. It is essential to measure the
      sensitivity of the detector efficiency to magnetic fields and
      show that the systematic effect introduced by the magnetic
      fields to the asymmetry is much smaller than the statistical
      error.
    \end{enumerate}
    
    There are a host of other issues to be addressed, including
    systematic effects, production and delivery of the polarized
    neutrons to the protons, the details of the neutron spin flipping,
    etc. However, the three issues mentioned above would render the
    experiment impossible if unresolved. Therefore we constructed an
    experiment to address these questions.

  \section{Experiment}
    \label{sec:prg}
      
    A proof-of-principle test run was performed at the pulsed
    spallation source of the Los Alamos Neutron Science Center
    (LANSCE) where neutrons are produced by impinging 800~MeV proton
    pulses at the repetition rate of 20~Hz from the Proton Storage
    Ring to the tungsten spallation target surrounded by a water
    moderator.  The neutron yield from the moderator surface has a
    Maxwell-Boltzmann low-energy component and a high-energy tail
    which falls off as $1/E$.  The peak of the neutron flux,
    $4\times10^{13}$~eV$^{-1}\cdot$s$^{-1}\cdot$sr$^{-1}$), is at
    about 4~meV \cite{Lis90}.  The intensity of the neutron flux at
    the moderator surface at epithermal energies is approximately
    given by
    \begin{equation}
      \frac{d^{2}N}{dt\,dE}\approx\frac{N_{0}f\Omega}{E}.
    \end{equation} 
    Here $N_{0}=2\times10^{12}$~s$^{-1}\cdot$sr$^{-1}$, $f$ is the
    fraction of the 13~cm by 13~cm moderator viewed by the collimator
    system, and $\Omega$ is the solid angle.  The experiment was
    mounted 6~m from the source and using the time-of-flight method
    the incident neutrons at an energy range below 1~eV were detected.
    This energy range was selected so that the 2.2~MeV gamma rates on
    the detector would correspond to the rates from the cold neutron
    moderator at the parity-violating experiment.  The neutrons were
    additionally thermalized in a polyethylene sample to take
    advantage of the larger capture cross section on a proton for the
    thermal neutrons.
     
\begin{figure}[htbp]
      \centering
      \epsfig{file=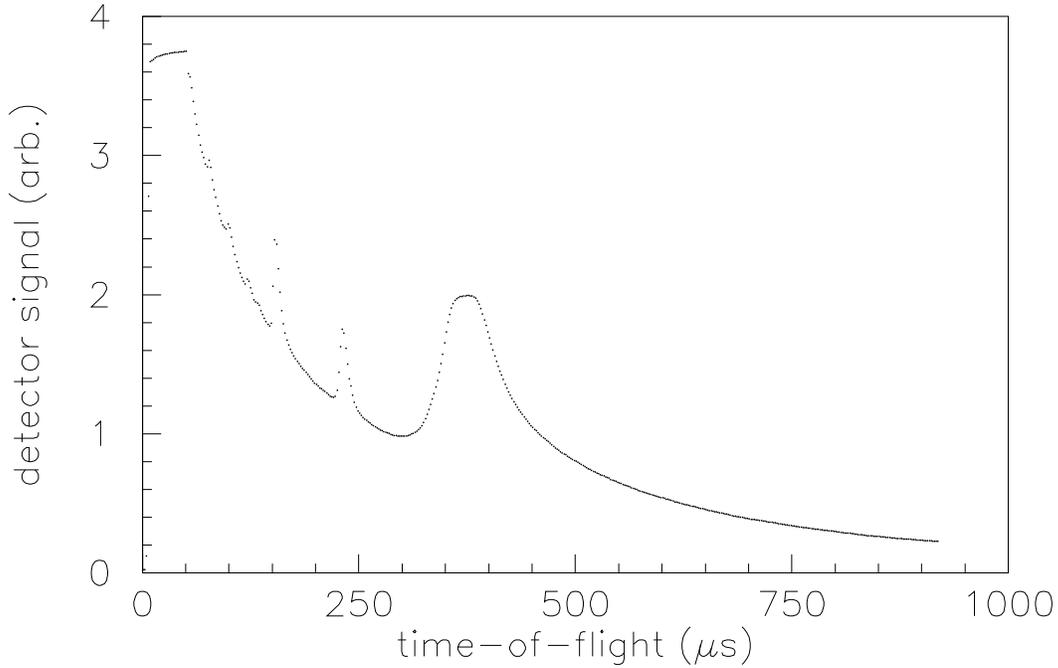,width=\textwidth}
      \caption{Gammas from neutron capture on polyethylene as a
        function of neutron time-of-flight. The neutron beam passes
        through 0.3~mm In and 1.5~mm Al before impinging on the
        target.}
      \label{fig:tof}
    \end{figure}

    Figure~\ref{fig:tof} shows the number of capture gammas from the
    polyethylene target as a function of neutron a time-of-flight
    (t.o.f.\ for 1~eV neutrons is about 440~$\mu$s at 6~m) after the
    beam has passed through a 0.3~mm thick indium foil and a 1.5~mm
    thick aluminum plate.  The signals were taken from the
    photocathodes of the photomultiplier tubes and then multiplied
    with preamplifiers.  The first goal of the test experiment was to
    verify that in the current-mode operation the electronic noise of
    the detector system can be reduced below the shot noise; the
    statistical fluctuations of the current due to the finiteness of
    the deposited energy of the gammas or the number of photoelectrons
    produced per event.  The second goal of the test experiment was to
    measure the sensitivity of the detector efficiencies to magnetic
    fields. For these tests an array of 12 CsI detectors were used.
    
\begin{figure}[htbp]
      \centering
      \epsfig{file=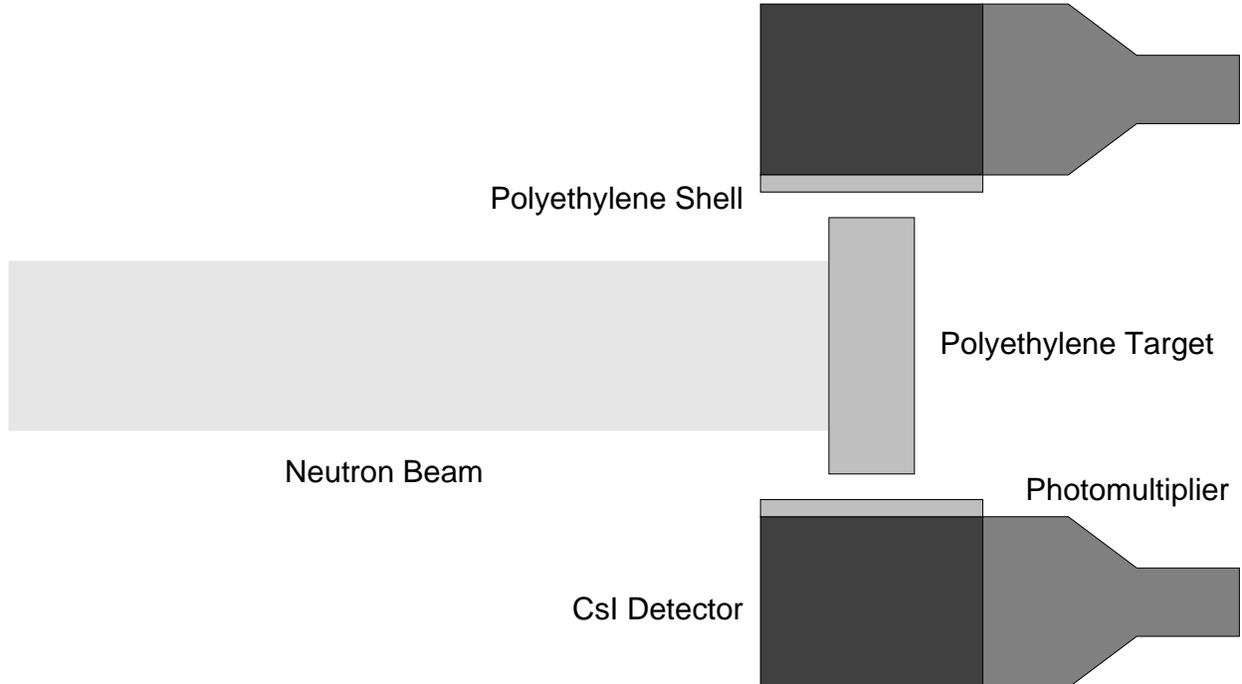,width=\textwidth}
      \caption{Experimental arrangement for the test
        experiment, showing the neutron beam, polyethylene target, and
        CsI detectors.}
      \label{fig:test}
    \end{figure}

    In the test experiment, a 15~cm diameter by 5~cm thick target of
    polyethylene, supported by a 2~cm thick cylindrical polyethylene
    shell, was placed in a 10~cm diameter neutron beam. The 12
    CsI(pure) crystals were arranged around the target in an annulus
    having an inner diameter of 20~cm, an outer diameter of 40~cm, and
    a length of 13~cm \cite{Fra94}. The detector array was centered
    approximately 6~m from the spallation neutron source and subtended
    a solid angle of 3.0~sr relative to the target. This arrangement
    is shown schematically in figure~\ref{fig:test}. The detector
    system was mounted inside a 10~cm thick lead housing with holes
    for the beam entry and exit.  The scintillation light from the CsI
    crystals was viewed directly by individual phototubes (Hamamatsu
    R5004). Signals were taken from the photocathodes. All other
    phototube elements, including dynodes, were connected to a battery
    and biased to $+90$~V with respect to the cathodes.
    
    Two different electronic configurations were tested. In the first,
    only two individual photocathodes were connected to low-noise
    preamplifiers attached directly to the phototube sockets. The
    amplified signals then went to a NIM module which formed sum and
    difference signals. This arrangement was used to measure the
    electronic noise and is similar to design for the proposed
    measurement, where each photocathode will have an individual
    preamplifier. In the second configuration, the detector
    photocathodes were connected together in two groups of six; top
    and bottom.  The two signals from each group were amplified and
    then they went to the summing and differential amplifiers, as
    described above. This circuit was used to study fluctuations in
    the capture gamma flux, due to effects such as beam intensity and
    position modulations. With this circuit, we could measure the
    spectral density of the beam noise.
    
\begin{figure}[htbp]
      \centering
      \epsfig{file=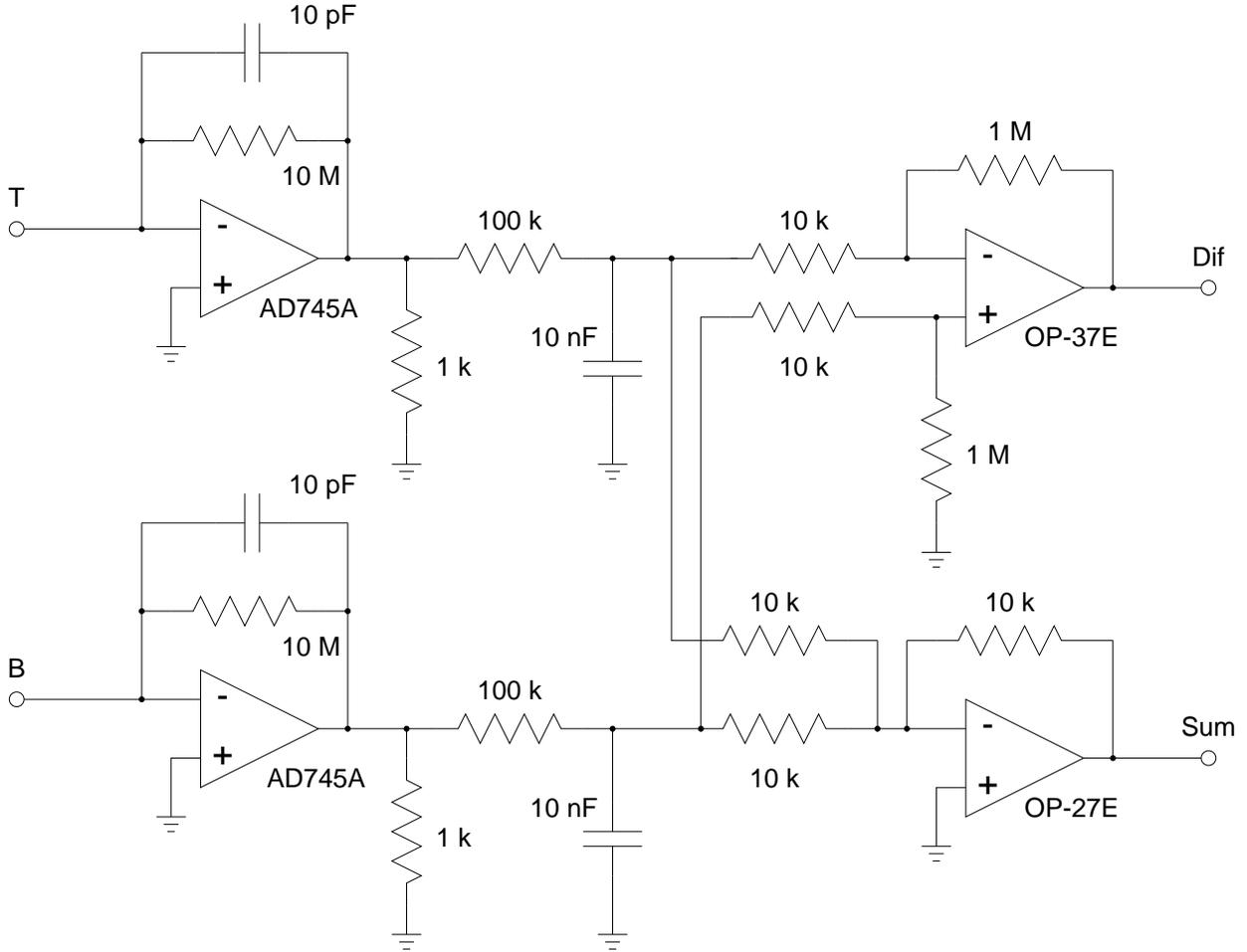,width=\textwidth}
      \caption{Diagram of the detector front-end electronics for the
        test experiment.}
      \label{fig:elec}
    \end{figure}

    The detector electronics used for the noise measurements are shown
    in figure~\ref{fig:elec}. The circuit consists of two low-noise
    current-to-voltage (IV) amplifiers, followed by stages which take
    the sum and difference of the two voltages.  The IV amplifiers
    were designed using AD745A op-amps. These op-amps have a typical
    voltage noise density of 2.9~nV/$\sqrt{\rm{Hz}}$ and typical
    current noise density of 6.9~fA/$\sqrt{\rm{Hz}}$ at 1~kHz.  The
    10~M$\Omega$ feedback resistor gives a DC gain of 10~V/$\mu$A\@. A
    10~pF feedback capacitor combined with an external low-pass filter
    limit the 3~dB bandwidth to 3.0~kHz. This corresponds to a time
    constant of approximately 50~$\mu$s.  Precision resistors (0.1\%)
    with low temperature coefficients (20~ppm/$^{\circ}$C) are used in
    the IV, summing, and differential amplifier stages. The summing
    circuit was designed to have an overall gain of 10~V/$\mu$A, and
    the differential circuit a gain of 1.0~V/nA\@. The spectral
    density measurement used the same circuit with the preamplifier
    gain reduced by a factor of 50 and the bandwidth increased to
    160~kHz.

    \subsection{Current Mode Signal}
      \label{sec:prg-cur}
  
      Several measurements were performed to ensure that the observed
      detector current for the full array, 0.56~$\mu$A for 1~eV
      neutrons, was consistent with calculations based on neutron
      capture gamma rays from hydrogen in the polyethylene target.
      This number is the product of the neutron flux, probability for
      the neutron to capture on a proton, probability that the capture
      produces a gamma ray which absorbs in the detector, and the
      charge per gamma ray which appears at the photocathode. The
      instantaneous flux of 1~eV neutrons on the target could not
      easily be measured directly.  Instead, a flux of
      $5\times10^{10}$~s$^{-1}$ was inferred from a measurement of the
      count rate in a $^{6}$Li-loaded glass scintillator 1~cm thick
      and 1~cm in diameter located 56~m from the spallation source.
      The observed count rate of 10.8~kHz was low enough to use the
      normal pulse-counting method.
  
      A Monte Carlo calculation was performed to estimate the fraction
      (approximately $60$\%) of incident neutrons which capture on
      protons in the polyethylene target after moderation.  Most of
      the remaining neutrons backscatter from the target. The mean
      free path of 5~mm for thermal neutrons in polyethylene was used
      in the calculations and was verified by relative transmission
      measurements. The prediction for the amount of backscattered
      neutrons was tested by moving the location of the target
      relative to the detector array and observing that the target
      position which gave the largest detector signal was located
      about 6~cm downstream from the center of the detector annulus.
      The large size of the signal at this position, about 2.5 times
      larger than the signal for the target centered on the array, was
      attributed to backscattered neutrons which capture on the
      protons in the cylindrical polyethylene support structure for
      the target (gamma rays from the support are closer to the
      detectors and have a larger solid angle).
  
      Since the solid angle was known, the only remaining quantity
      needed to calculate for the detector current was the number of
      photoelectrons produced in the photocathodes of the CsI
      detectors by the 2.23~MeV gamma ray. This number was determined
      by two independent methods.  First, the polyethylene target was
      replaced by a 0.3~mm thick In foil. At 1.46~eV indium possesses
      a strong neutron resonance which captures all the neutrons from
      the beam.  The resonance state decays primarily \emph{via} gamma
      emission. Using the measured detector current at the resonance,
      we inferred a value of 70 photoelectrons per MeV of deposited
      gamma ray energy. This number was consistent within errors with
      the value of 65 photoelectrons per MeV inferred from the energy
      resolution measurement of the CsI detectors with a $^{60}$Co
      gamma source.  The number of photoelectrons per MeV was
      calculated under the assumption that the energy resolution is
      dominated by the statistical fluctuations of the number of
      photoelectrons.
  
      Based on this combination of measurements and simulations, the
      predicted detector current for the full array of 12 detectors is
      about $0.5\pm0.1$~$\mu$A for 1~eV neutron rate. The agreement
      with the measured value of 0.56~$\mu$A has two consequences: 1)
      the current-mode signal is indeed dominated by gamma rays from
      neutron capture on protons, 2) the value for the number of
      photoelectrons per MeV in the detector is determined.  This
      value is needed to calculate the expected amount of noise in the
      detector array due to current shot noise from neutron counting
      statistics.

    \subsection{Electronic Noise}
      \label{sec:prg-noi}
  
      Using a digital oscilloscope (LeCroy model TDS 744A) the
      electronic noise of the detector was determined for the sum and
      difference outputs. The noise of the sum output was
      760~$\mu$V$_{\rm{rms}}$\@. Referred to the input, this
      corresponds to 1.4~pA/$\sqrt{\rm{Hz}}$. The difference output
      had 110~$\mu$V$_{\rm{rms}}$ noise, corresponding to
      0.2~pA/$\sqrt{\rm{Hz}}$ at the input.  It is useful to compare
      the difference noise to the rms shot noise created by the
      quantized nature of the expected detector signal
      \begin{equation}
        I_{sn}/\sqrt{f}=\sqrt{2qI},
      \end{equation}
      where $q$ is the charge generated in the photocathode per gamma
      and $I$ is the photocathode current. Using the expected values
      $q\approx150e$ and $I\approx100$~nA, we obtain $I_{sn}/\sqrt{f}
      \approx$~2~pA/$\sqrt{\rm{Hz}}$, a factor of 10 greater than our
      electronic noise. Calculations with a SPICE model of the
      preamplifier indicate that an electronic noise of just
      13~fA/$\sqrt{\rm{Hz}}$ should be obtainable. We attribute the
      excess noise observed in our test to be from electromagnetic
      pickup.  This contribution will be reduced with improved
      electrostatic shielding.

    \subsection{Spectral Density}
      \label{sec:prg-den}
      
      Periodic variations in experimental parameters can produce false
      asymmetries. These effects include fluctuations in detector
      gain, beam intensity and beam position, which are expected to
      dominate over other fluctuations. The proposed experiment has
      been designed so that these fluctuations do not contribute to a
      false asymmetry in first order. Since the parity-violating
      asymmetry is formed by taking the difference between up and down
      detector currents, the two currents are measured simultaneously,
      making the apparatus quite insensitive to incident flux
      variations. Second, the azimuthal symmetry of the apparatus
      suppresses the contribution from the beam motion.  While such
      effects do not lead directly to false asymmetries, it is
      possible for them to contribute in higher order.  For example,
      the beam motion combined with detector-efficiency differences
      can give a non-zero effect. For this reason, it is important to
      know the size of the beam fluctuations.
  
\begin{figure}[htbp]
        \centering
        \epsfig{file=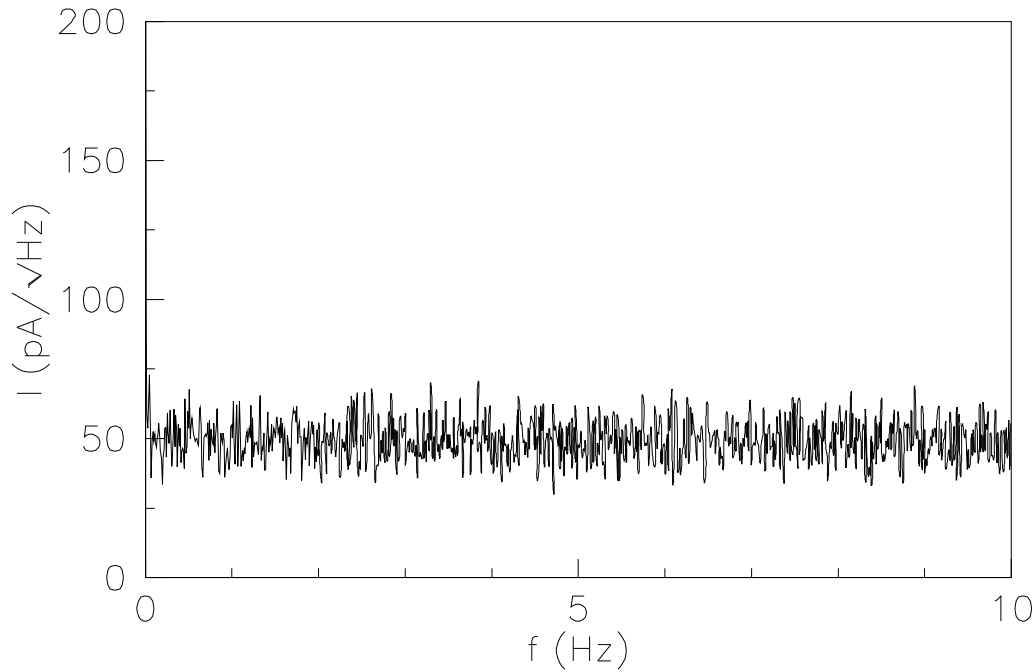,width=\textwidth}
        \caption{Spectral density of the detector difference signal,
          due to fluctuations in the neutron beam parameters.}
        \label{fig:spec}
      \end{figure}

      We measured the influence of these fluctuations on the detector
      signal and set an upper limit for their contribution. We sampled
      the difference signal from the detectors every 50~ms (every beam
      pulse) by integrating the voltage for $1\mu$s for a 20~minute
      period while the spallation source operated in a steady-state
      mode. The autocorrelation function of the difference signal,
      $f(\tau)=\langle i(t)i(t-\tau)\rangle$ was then calculated. The
      Fourier transform of this quantity,
      $F(\omega)=\int^{+\infty}_{-\infty}f(\tau)e^{i\omega\tau}d\tau$,
      is the spectral density of the intensity fluctuations of the
      neutron source as filtered through the difference signal.  There
      were no periodic sources of noise observed in the 0--10~Hz range
      (fig.~\ref{fig:spec}). The upper limit of this range is set by
      the sampling rate, which is in turn limited by the pulse
      spacing.  Because the circuit bandwidth (160~kHz) was larger
      than this upper observable frequency, the noise above 10~Hz is
      aliased into the 0--10~Hz range.  Correcting for this effect, we
      estimate an upper limit of 5~fA/$\sqrt{\rm{Hz}}$ on beam-induced
      fluctuations on the difference signal.

    \subsection{Detector Sensitivity to Magnetic Fields}
      \label{sec:prg-mag}
  
      The primary techniques for reducing the false asymmetries
      generated by the experimental fluctuations is the fast reversal
      of the neutron spin.  Depending on the reversal techniques it
      will introduce changing magnetic fields which can effect the
      efficiency of the detector. Therefore it was important to
      measure the sensitivity of the detector to magnetic fields.  The
      measurement was performed by applying a 10~gauss magnetic field
      to the detectors using a pair of 76~cm diameter coils. The field
      direction was reversed every 10~s. The field was oriented
      parallel to the photocathode surfaces of the photomultiplier
      tubes in an attempt to maximize the size of the effect. The
      observed change in the detector efficiency was $2\times10^{-5}$
      per gauss. This is about five orders of magnitude smaller than
      one would expect for a typical photomultiplier tube operated in
      the normal fashion with high voltage on the dynodes and the
      signal measured from the anode.  The observed field sensitivity
      can be reduced further through the use of special photocathode
      materials.
      
      Changes in static magnetic fields at the detectors upon a spin
      flip must therefore be held below the 10~$\mu$gauss level.  Spin
      flipping by RF magnetic fields would be preferable, since one
      expects the detectors and associated electronics to be even less
      sensitive to such fields and one can shield RF magnetic fields
      effectively.
      
      The conclusions of the test measurements are that the noise
      introduced by the current-mode detection is less than the shot
      noise and the magnetic field sensitivity of the detector is
      very low when the signals are taken from the photocathodes.
      
      In the remainder of the paper, we discuss a conceptual design of
      the experiment based on the results of the test experiment, an
      estimate for the statistical accuracy, and diagnostics for some
      classes of systematic effects.

  \section{Conceptual Design of $\vec{n}+p\rightarrow d+\gamma$ Experiment}
    \label{sec:des}
    
\begin{figure}
      \centering
      \epsfig{file=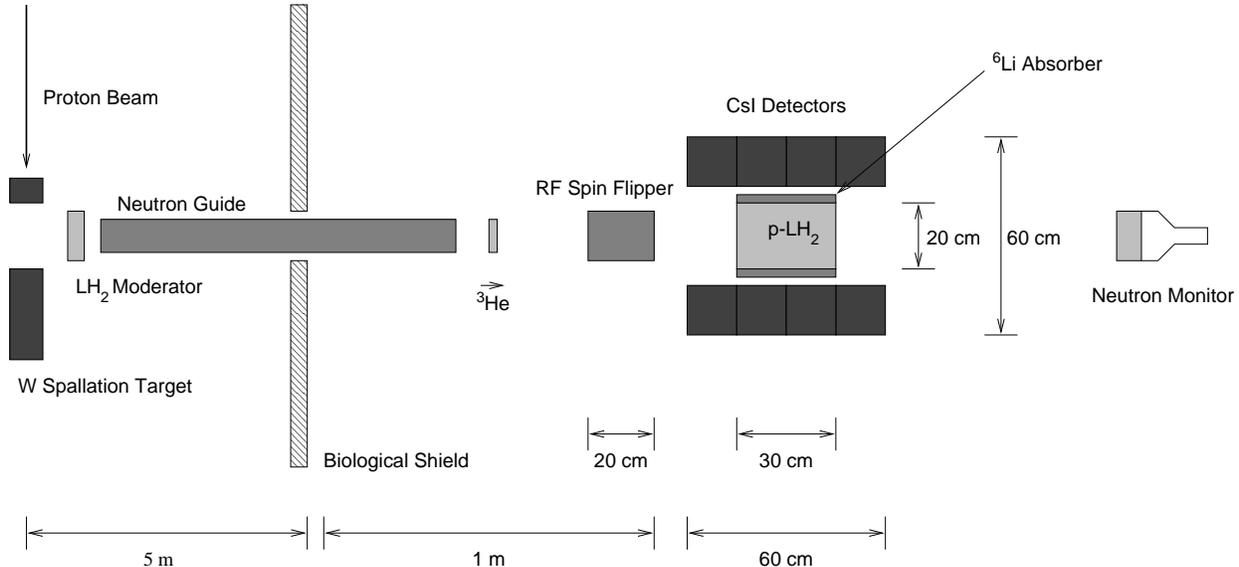,width=\textwidth}
      \caption{The conceptual design for the $\vec{n}+p\rightarrow d+\gamma$ 
        experiment, showing the most important elements (not to
        scale). Approximate sizes and distances are indicated for some
        features.}
      \label{fig:dsgn}
    \end{figure}

    We briefly describe in this section a conceptual design of the
    $\vec{n}+p\rightarrow d+\gamma$ experiment.  Figure~\ref{fig:dsgn}
    shows the elements of the design. Next we discuss those aspects
    which are particularly important for the experiment.
      
\begin{figure}
      \centering
      \epsfig{file=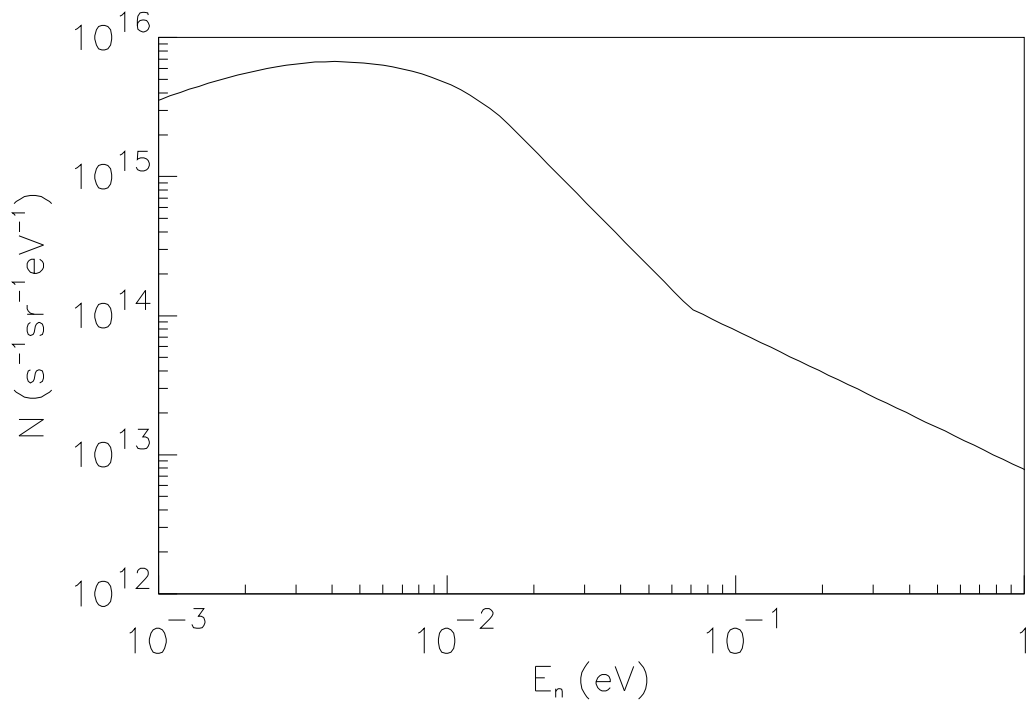,width=\textwidth}
      \caption{Neutron flux from the LANSCE coupled cold
        moderator. This is the total neutron flux from the 13~cm by
        13~cm moderator surface with the average proton current of
        200~$\mu$A\@.}
      \label{fig:flux}
    \end{figure}

    The experiment requires a high flux of cold neutrons with energies
    below 15~meV, as will be explained later. While such neutrons are
    available from cold moderators at both reactors and spallation
    neutron sources, the pulsed nature of the neutron flux from a
    pulsed spallation source provides a very powerful diagnostic tool
    for a number of systematic effects for this experiment. At LANSCE
    the cold neutron source consists of a liquid hydrogen moderator
    coupled to the 20~Hz pulsed neutron source. The neutron spectrum
    consists of a Maxwellian component with a maximum at 4~meV and
    width set by the effective temperature of the moderator
    (approximately 50~K) and a higher energy component of
    ``under-moderated'' neutrons with an approximately $1/E$ spectrum
    (figure~\ref{fig:flux}). Approximately 90\% of the neutrons
    possess energies below 15~meV \cite{Fer97}.
    
    The intensity of neutrons from an isotropic source falls off as
    $1/R^{2}$. At cold neutron energies, however, it is possible to
    use neutron guides to transport neutrons.  The neutron guide
    possess a reflectivity which is close to unity for neutrons
    incident at glancing angles below a critical angle $\theta_{c}$,
    and the reflectivity falls sharply above this angle. The function
    of the neutron guide is to conserve the high neutron flux
    available near the moderator.  Since the $\theta_{c}$ depends on
    neutron energy, the guide increases the flux at lower neutron
    energies compared to the neutron transmission without the guide.
    The experiment requires polarized neutrons.
    
\begin{figure}
      \centering
      \epsfig{file=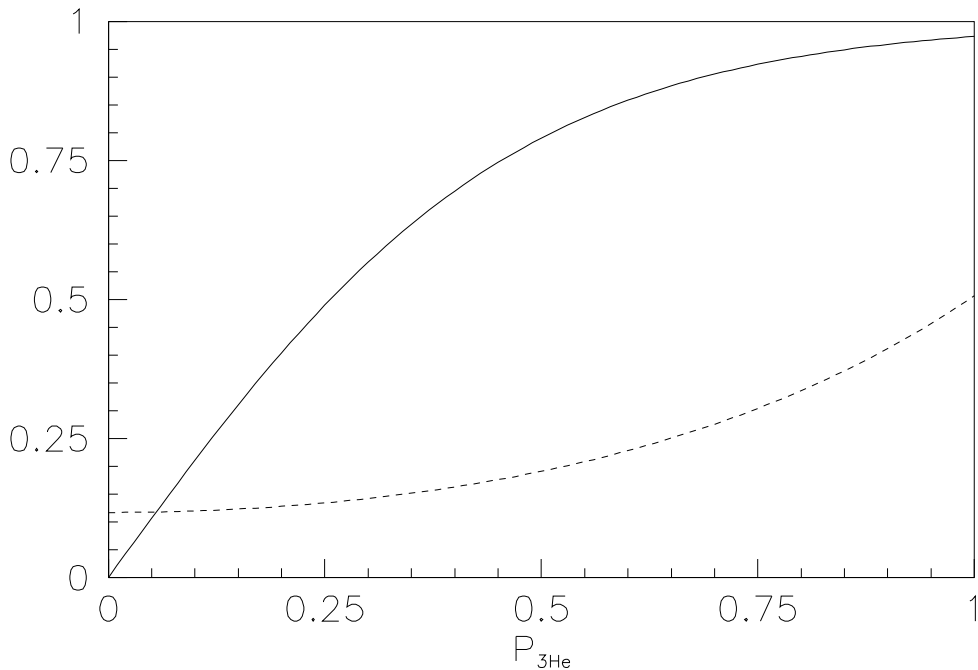,width=\textwidth}
      \caption{Polarization (solid) and transmission (dash) of 4~meV
        neutrons after passage through a 6~atm$\cdot$cm $^{3}$He cell,
        as a function of $^{3}$He polarization.}
      \label{fig:hepol}
    \end{figure}

    We plan to use a polarized $^{3}$He neutron-spin filter
    \cite{Cou90,Tas92}.  Polarized $^{3}$He gas acts as a transmission
    polarizer. The total cross section for the $^{3}$He nucleus is
    very large for the $J=0$ capture channel, which decays mostly by
    breakup to a triton and proton, and about four orders of magnitude
    smaller for the $J=1$ channel. The neutron polarization ($P$) and
    transmission ($T$) properties are therefore strongly dependent on
    $^{3}$He polarization: an example is shown in
    figure~\ref{fig:hepol}.
      
    The parity-violating signal will be isolated by periodically
    reversing the neutron spin. The spin flipper must be capable of
    rapidly and reproducibly reversing the beam polarization in a
    broad range of cold neutron energies with essentially unit
    efficiency over the entire beam profile with no effect on the
    gamma detectors.  The fast (20~Hz) spin reversal is possible with
    a spin flipper which uses RF magnetic fields \cite{Gri97}.  The
    reversal of the neutron polarization then consists of turning the
    RF magnetic field on and off, which can be done quickly enough to
    flip the neutron polarization on every beam burst at 20~Hz.  The
    gamma detectors can be efficiently shielded from the RF magnetic
    field with a thin metal enclosure.  Neutron spin flippers based on
    the reversal of an otherwise static magnetic field suffer
    primarily from the greater difficulties of shielding static
    magnetic fields and the relatively higher sensitivity of the gamma
    detector efficiency to static fields.
      
    In order for the experiment to be clearly interpretable as a
    measurement of parity violation in the neutron-proton system, a
    target of pure hydrogen is required. Clearly, it is essential that
    the polarized neutrons retain their polarization until they
    capture. It is therefore important to consider the spin dependence
    of the scattering.
      
    The ground state, the para state, of the hydrogen molecule has
    $J=L=S=0$, and the first excited state, the lowest ortho state, is
    at 15~meV above the para state.  A large fraction of the cold
    neutrons possess energies lower than 15~meV\@. Since these
    neutrons cannot excite the para-hydrogen molecule, only elastic
    scattering and capture are allowed, and spin-flip scattering is
    forbidden. The neutron polarization therefore survives the large
    number of scattering events which occur before the capture. Higher
    energy neutrons will undergo spin-flip scattering and will
    therefore depolarize. This is the reason for the requirement of
    neutrons with energies below 15~meV\@.
      
    We must prepare a liquid hydrogen target in the para state.  For
    liquid hydrogen held at 20~K and atmospheric pressure the
    equilibrium concentration of para-hydrogen is 99.8\%, low enough
    to ensure a negligible population of ortho-hydrogen.
    
    Finally, we must detect the 2.2~MeV gamma rays from neutron
    capture. Given the small size of the expected asymmetry and
    gamma-detector dead times, the required counting rates for a
    practical experiment are too great for pulse counting.
    Current-mode gamma detection is therefore required.  In addition,
    the gamma detector must cover a large solid angle with a high,
    time-independent efficiency which is unaffected by neutron spin
    reversal and radiation damage.
      
    Segmentation of the detector is required to resolve the angular
    dependence of the expected parity-violating signal and
    discriminate false effects. For example, there is a predicted
    parity-conserving gamma asymmetry of
    $A^{PC}_{\gamma}=7\times10^{-9}$ \cite{Cso97} in the \emph{np}
    capture.  This process gives an asymmetry about the same magnitude
    as our sensitivity goal for the parity violating signal, but with
    an angular distribution orthogonal to the parity violating signal.
    Some degree of segmentation is therefore necessary to separate the
    parity-violating and parity-conserving asymmetries.
      
    The alkali iodine crystals are well suited for the gamma ray
    detection because of their high density, 4.5~g/cm$^{3}$.  Three
    interaction lengths ($3\times5.5$~cm) thick crystal will stop
    95~\% of the 2.2~MeV gamma rays. Our plan is to use cubes of
    CsI(Tl) crystals approximately 15~cm on a side coupled to
    photomultiplier tubes operated in current mode. The signal is
    taken directly from the photocathode to decrease the sensitivity
    of detector gain to magnetic fields, as in the test experiment.
    Similar detectors have been observed to have acceptably low gain
    drifts ($<1$\% per week \cite{Frl97}). Based on previous studies,
    these detectors should be capable of withstanding the integrated
    gamma dose in the experiment without serious effects on the
    detector efficiency \cite{Wei93}. They were used successfully in
    our test runs to perform current-mode counting at the expected
    instantaneous rates of $10^{11}$ gammas per second.
      
    The gamma detectors must be shielded from neutrons scattered from
    the hydrogen target. $^{6}$Li is the ideal material for this
    purpose due to its large neutron absorption cross section, most of
    which proceeds by a breakup channel to a triton and alpha in their
    ground states, thus producing no gamma rays.
      
\begin{figure}
      \centering
      \epsfig{file=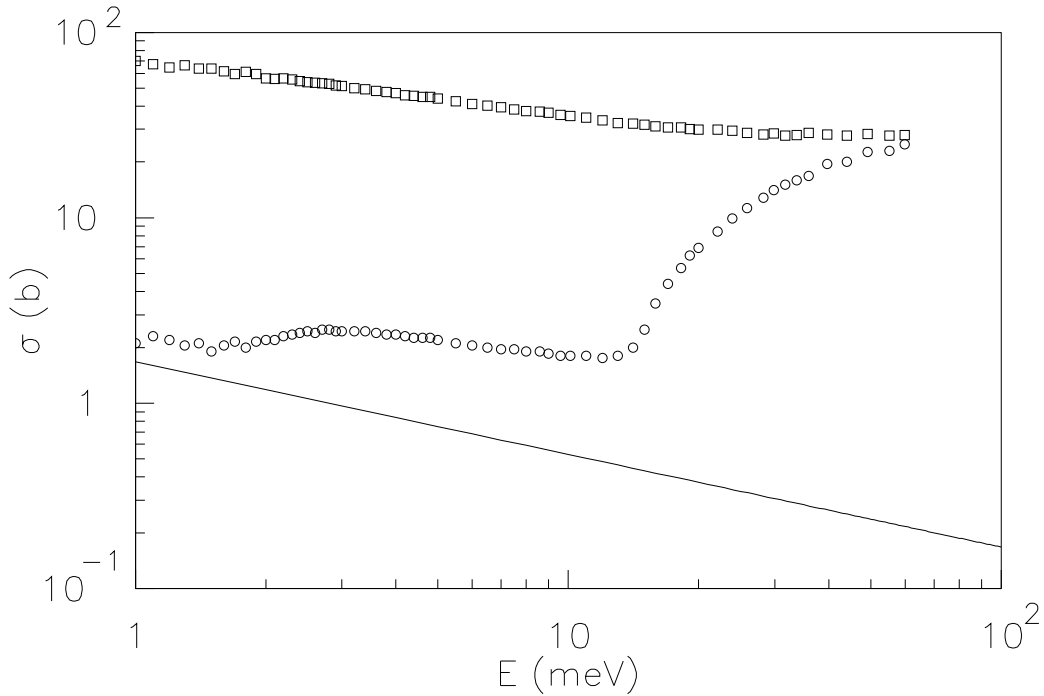,width=\textwidth}
      \caption{Measured scattering cross sections for neutrons on 
        para-hydrogen (solid) and neutrons on normal hydrogen (dash).
        The solid line is the cross section for the \emph{np}
        capture.}
      \label{fig:oph}
    \end{figure}

    In addition to its primary purpose of measuring the beam
    intensity, the beam monitor can be used to detect changes in the
    concentrations of ortho-hydrogen and para-hydrogen in the target.
    This is possible due to the very large difference in neutron cross
    sections for the two species at low neutron energies (see
    Figure~\ref{fig:oph}) \cite{Sei70a,Sei70b}.  At 4~meV neutron
    energy, for example, the ortho-hydrogen cross section is 20 times
    larger than the para-hydrogen cross section.  Even small changes
    in the 0.2\% equilibrium ortho-hydrogen concentration in the
    target will therefore lead to large changes in the transmitted
    neutron flux.
    
    Next we estimate briefly the statistical accuracy of the
    experiment achievable with the LANSCE neutron sources described
    above.  The estimate has been done under the following
    assumptions:
    \begin{enumerate}
    \item The time-averaged brightness from a coupled liquid hydrogen
      moderator at the (upgraded) LANSCE proton beam current of
      200~$\mu$A peaks at 4~meV at a value
      $6\times10^{15}$~neutrons/(eV$\cdot$s$\cdot$sr) (see
      figure~\ref{fig:flux}).
    \item A 5~m, 10~cm$\times$10~cm multi-layer neutron guide is
      mounted 1~m from the moderator. The transmission of the guide is
      assumed to be 100\% for neutrons with angles of incidence with
      the guide surface smaller than $2\theta_{c}$.
    \item A polarized $^{3}$He neutron polarizer with 65\% $^{3}$He
      polarization and 4.5~atm$\cdot$cm of gas.
    \item A 30~cm long, 30~cm diameter cylindrical liquid
      para-hydrogen target, which converts approximately 60\% of the
      incident neutrons to gamma rays according to MCNP calculations.
    \item The CsI(Tl) gamma detectors are used only to determine which
      hemisphere the gamma ray enters.
    \item The gamma detectors cover $4\pi$ solid angle.
    \item The uncertainty in the gamma flux striking detectors,
      operated in current mode, is dominated by gamma ray counting
      statistics.
    \end{enumerate}
      
    Under these assumptions, a Monte Carlo simulation gives a
    statistical error in the gamma asymmetry in one year of counting
    of $5.8\times10^{-9}$. This statistical accuracy suffices to reach
    the goal of the experiment.

  \section{About Systematic Uncertainties}
    \label{secsys}
    
    The most challenging aspect of the experiment is to design an
    apparatus which is a) insensitive to spurious systematic effects,
    b) incorporates a large number of independent methods to isolate a
    true parity-violating signal, and c) is flexible enough to allow
    the size of as many systematic effects as possible to be magnified
    artificially. The most dangerous class of systematic effects are
    those correlated with the neutron spin reversal, either through
    the direct physical effects due to the interaction of the neutron
    spin itself or through indirect effects on the gamma ray detection
    efficiency correlated with the neutron spin flipping process.
    
    A detailed analysis of systematic effects lies outside the scope
    of this paper.  However, we can point to aspects of the design of
    the experiment which are important for the study and isolation of
    systematic effects.
   
    At a pulsed neutron source, the arrival time of the neutron at the
    experiment after the proton pulse strikes the spallation target
    depends on the neutron energy: $t\propto1/\sqrt{E}$.  The relation
    between the neutron energy and the timing of the gamma ray signal
    is blurred somewhat by the distribution of moderation times of the
    neutrons first in the LH$_{2}$ moderator of the spallation source
    and then in the hydrogen target (which averages about 100~$\mu$s),
    but this effect is small in comparison to the flight times of the
    neutrons of interest (for instance, a 4~meV neutron needs 9~ms for
    a 8~m flight).  Therefore, the time dependence of the gamma ray
    signal can be correlated with the incoming neutron energies.
  
    The time-of-flight information is very useful for isolating and
    identifying systematic effects simply because different effects
    possess different dependences on neutron energy. For example, a
    systematic effect associated with the neutron beam motion due to
    the Stern-Gerlach effect in a gradient magnetic field in the spin
    transport system grows as the square of the time spent in the
    field gradient, and thus increases as a function of
    time-of-flight. On the other hand, a systematic effect from
    parity-conserving left-right scattering asymmetries coupled to
    detector asymmetries grows with the incident neutron energy and
    therefore decreases with time-of-flight. Each systematic effect
    has a characteristic time signature which can be used as a partial
    means of identification. Most of these time signatures are
    different from that expected from the true signal.
    
    As noted in section~\ref{sec:des}, neutrons with energies above
    15~meV tend to depolarize before they capture. This not only
    effectively turns off the parity violating signal, but also all
    systematic effects which require the neutrons to be polarized in
    the para-hydrogen target. For example, effects induced by
    left-right scattering asymmetries in para-hydrogen must vanish
    above 15~meV\@. On the other hand, systematic effects not
    associated with polarized neutrons on hydrogen, such as
    bremsstrahlung from the parity-violating beta decay of window
    materials of the LH$_{2}$ target cryostat, will still be present
    at neutron energies above 15~meV\@.

  \section{Conclusions and Summary}
    \label{sec:con}
 
    A sensitive measurement of the parity-violating gamma asymmetry in
    the reaction $\vec{n}+p\rightarrow d+\gamma$ can give definitive
    information on one of the most important and interesting
    components of the weak \emph{NN} interaction.  We have
    successfully tested a current-mode detector concept for the gamma
    rays and verified that its precision is limited only by gamma ray
    counting statistics. We have demonstrated that the efficiency of
    our current-mode gamma detector is insensitive to changes in
    magnetic field, the most important external parameter used to
    change the sign of the parity-violating signal in the experiment.
    The realization of all other aspects of the experimental design
    appears to pose no insuperable technical difficulties.  We have
    discussed a realistic experimental design which has the potential
    to be free of systematic effects at the required level and
    incorporates a number of powerful diagnostics to isolate
    systematic effects.
    
    We conclude that an experiment to search for the parity-violating
    gamma asymmetry in the reaction $\vec{n}+p\rightarrow d+\gamma$
    with a sensitivity which is likely to obtain a nonzero result is
    now feasible.
 
  \section{Acknowledgments}
    \label{sec:ack}
  
    This work was supported in part by the U.S. Department of Energy.

\end{document}